\documentstyle[emulateapj]{article}
\slugcomment{Accepted for publication in {\it Astrophysical Journal}, part 2, July 9, 1999}

\begin{document}

\title{PSR 0943+10: a bare strange star?} 
\author{R.X.Xu\altaffilmark{1,2}, G.J.Qiao\altaffilmark{4,1,2},
 Bing Zhang\altaffilmark{3}}
\altaffiltext{1}{Beijing Astronomical Observatory and National Astronomical
 Observatories, Chinese Academy of Science (CAS), Beijing 100012, P.R. China}
\altaffiltext{2}{CAS-PKU Beijing Astrophysics Center and Astronomy
 Department, Peking University (PKU), Beijing 100871, P.R. China.
 emails: RXXU@bac.pku,edu.cn, gjn@pku.edu.cn}
\altaffiltext{3}{Laboratory of High Energy Astrophysics, NASA Goddard Space
 Flight Center, Greenbelt, MD 20771. National Research Council Research
 Associate. email:bzhang@twinkie.gsfc.nasa.gov}
\altaffiltext{4}{China Center of Advanced Science and Technology
 (World Laboratory), P.O. Box 8730, Beijing 100080, P.R. China}

\begin{abstract}

Recent work by Rankin \& Deshpande strongly suggests that there exist strong 
``micro-storms'' rotating around the magnetic axis of the 1.1s pulsar PSR 0943+10. 
Such a feature hints that most probably the large-voltage vacuum gap proposed by 
Ruderman \& Sutherland (RS) does exist in the pulsar polar cap. However, there are
severe arguments against the formation of the RS-type gap in pulsars, since the
binding energies of both the $^{56}$Fe ions and the electrons in a neutron star's 
surface layer is too small to prevent thermionic ejection of the particles from the
surface. 
Here we propose that PSR 0943+10 (probably also most of the other 
``drifting'' pulsars) might be bare strange stars rather than normal neutron stars, 
in which the ``binding energy'' at the surface is merely infinity either for the 
case of ``pulsar'' or ``anti-pulsar''. It is further proposed that identifying a 
drifting pulsar as an anti-pulsar is the key criterion to distinguish strange 
stars from neutron stars.

\keywords{pulsars: general -- pulsars: individual: PSR 0943+10 -- stars: neutron 
--  elementary particles}
\end{abstract}

\section{Introduction}

A wealth of observations has been collected for pulsars since the discovery
of them more than thirty years ago. However, some important discrepancies 
still remain in our understanding of the particle acceleration mechanisms, the 
emission processes, and even the nature of pulsars. It is commonly agreed 
that there exists an inner accelerator near the magnetic polar cap region of 
a pulsar, but two subclasses of models appear in the literatures. The 
space-charge-limited flow models (Sturrock 1971; Arons \& Scharlemann 1979;
Arons 1983; Muslimov \& Tsygan 1992; Harding \& Muslimov 1998) assume free
ejection of particles of either sign from the star surface. Another type
of models, however, assumes that certain particles (usually ions) could
be bound in the surface layer of the star, so that a vacuum gap can form
and keep breaking down to generate ``sparks'' continuously (Ruderman \&
Sutherland 1975, hereafter RS75; Usov \& Melrose 1995, 1996, hereafter UM95,
UM96; Zhang \& Qiao 1996; Zhang et al. 1997). Both subclasses of models have 
some observational supports, and it is very likely, different kinds of 
accelerators may exist in different pulsars.
 
Maybe the strongest observational support to the RS-type vacuum gap model
is the regular ``drifting'' of the subpulses observed from some pulsars. In 
such a model, the sparks produced by inner gap breakdown provide the source
of the subpulses, and ${\bf E\times B}$ drift due to the lack of charges
within the gap causes the observed ``drifting'' phenomena. In their original
paper (RS75), Ruderman \& Sutherland has treated the ``drifting sparking'' 
process carefully in a detailed calculable way to get predicted value of $P_3$ 
to be directly comparable with the observations. This was not done by any other 
models hitherto known. 

However, the RS model encounters severe theoretical criticisms known as the 
so-called ``binding energy problem''. Even if it has been modified by different 
ways either by introducing partial screening of the parallel electric fields 
(UM95; UM96) or by reducing gap height with inverse Compton scattering-induced 
breakdown (Zhang \& Qiao 1996; Zhang et al. 1997), the formation of such 
vacuum gaps is still suspected, since calculations using various methods show 
that the binding energy of the $^{56}$Fe ions is much lower than what is required 
to maintain a vacuum gap or simply there is no binding at all (e.g. Flowers et al. 
1977; M\"uller 1984; Jones 1986; Neuhauser, Koonin, \& Langanke 1987; K\"ossl et 
al. 1988). Furthermore, The RS model is only viable for the case of anti-parallel 
rotator, i.e., ${\bf \Omega}\cdot {\bf B} < 0$, which they defined as ``pulsars''. 
In such a geometry, positive charges are expected to dwell on the polar cap, and 
a vacuum gap could be formed if positive ions {\em could} be bound within the 
molecular lattice of the star layer. For a parallel rotator with ${\bf \Omega}
\cdot {\bf B} > 0$, however, copious negative electrons could flow freely out 
from the surface, so that a vacuum gap will never form. This is referred to as 
``antipulsars'' in the RS's term.

Recently Deshpande and Rankin have developed a technique for ``mapping'' the
pattern of polar cap sparks or ``micro-storms''. They studied the typical
drifting pulsar PSR 0943+10 applying this technique and came to a clear map
of the polar cap sparking pattern of this pulsar (Rankin \& Deshpande 1998,
Deshpande \& Rankin 1999; for popular report, see Glanz 
1999). Their results strongly suggest that the RS vacuum gap does exist in this
pulsar's polar cap. Vivekanand \& Joshi (1999) also independently came to the 
similar conclusion by studying the competing drifting subpulses in PSR 0031-07, 
and argued that ``there is a genuine need to reinvestigate the theoretical
basis of this model'' (RS model). All these again pose the question how 
certain charged particles can be bound in the star surface. The question 
becomes more severe if the drifting pulsar can be identified as an 
``anti-pulsar''\footnote{PSR 0943+10 was reported to be an anti-pulsar in 
Rankin \& Deshpande (1998), but this conclusion is not solid (Rankin 1999, 
personal communication).}. 

On the other hand, another kind of astrophysical objects, the strange stars,
have been widely discussed (e.g. Witten 1984; Alcock, Farhi, \& Olinto 1986).
Trying to find criteria to distinguish strange stars from neutron stars not 
only is an interesting topic in astrophysics, but will, also exert important
impact on fundamental physics. However, there seems to be no evident criteria
since strange stars are analogous to neutron stars in many aspects (for recent
review, see Lu 1998). Specifically, the canonical strange star models
all invoke a solid crust which is composed of normal matters (Alcock, Farhi, 
\& Olinto 1986; Glendenning \& Weber 1992; Huang \& Lu 1997). This makes 
strange stars indistinguishable from neutron stars in appearance. Hence, 
all the difficulties faced by the RS vacuum gap model for neutron stars 
still remain. 

Recently, Xu \& Qiao (1998, hereafter XQ98) proposed that a magnetosphere 
similar to that of a neutron star could also be formed outside a bare strange 
star, so that bare strange stars can act as pulsars as well. Here we'll show that 
an RS-type vacuum gap can be well formed above the surface of a bare strange star 
both for the case of ${\bf \Omega}\cdot {\bf B} < 0$ {\em and} the case of ${\bf 
\Omega}\cdot {\bf B} > 0$ (anti-pulsar). We suggest that PSR 0943+10 and other 
drifting pulsars might be bare strange stars rather than normal neutron stars.


\section{Binding energy problem in RS model for neutron stars or strange stars 
with crusts}

The essential condition of the RS vacuum gap model is that certain charged 
particles could be bound in the star surface, so that a boundary condition
of ${\bf E \cdot B}\neq 0$ is satisfied at the surface. In their original work
(RS75), Ruderman \& Sutherland have set two criteria to judge whether ions
are bound in the surface. These two criteria were later more evidently
presented by Usov \& Melrose (UM95): 1. whether thermionic emission is 
important: i.e., whether the surface temperature is in excess of a critical
``unbound'' temperature; or 2. whether field emission is important: i.e., whether
the parallel electric field is in excess of a critical ``unbound'' field value.
Both the critical temperatures and the critical electric fields are determined
by the work functions of the particles. For electrons, the work function is
just their Fermi energy, which reads
$w_{\rm e,NS}=\epsilon_{\rm F}={2\pi^4\hbar^4 
c^2 \over e^2 B^2 m_{\rm e}} n_{\rm e}^2\simeq 1.03\times 10^{-63}  B_{12}^{-2} n_{\rm e}^2$, 
where $n_{\rm e}$ is the electron number density (Ruderman 1971; Flower et al. 1977; 
UM95). For $^{56}$Fe ions, the work function is the cohesive energy per ion in 
the {\em assumed} magnetic metal, which is quite uncertain, since this small 
number is the difference of two large numbers (e.g. M\"uller 1984). UM95 has 
adopted $w_{\rm i,NS}=\Delta\epsilon_{\rm c} \simeq (0.9{\rm keV}) B_{12}^{0.73}$
following Abrahams \& Shapiro (1991). However, the error of the calculation
is of the order of this value itself. Thus it is possible that there
might be no cohesive energy at all.

The critical temperature for thermionic emission to be important is calculated
by equating the current density due to thermionic emission (which is $\propto \exp
(-w/kT)$, see eq.[12] of RS75 and eq.[2.10] of UM95) with the Goldreich-Julian 
(1969) charge density $n_{\rm GJ}ec$, where $n_{\rm GJ}\simeq \Omega B/(2\pi ce)$. 
For the case of a parallel rotator (${\bf \Omega}\cdot {\bf B}>0$, i.e. anti-pulsar), 
electrons are expected to be pulled out from the surface. Adopting the density at
surface as $\rho \simeq (4\times 10^3 {\rm g~cm}^{-3}) B_{12}^{6/5} A_{56} Z_{26}
^{-3/5}$ and noticing $n_{\rm e}=\rho/m_{\rm p} \left({Z\over A}\right)$,
the work function of 
electrons is then $\sim 0.8~$keV. Thus the critical temperature for electron thermion 
ejection is $T_{\rm cri,e}\simeq (3.7\times 10^5{\rm K}) Z_{26}^{4/5} B_{12}^{2/5}$ 
(UM95; UM96). 
For an anti-parallel rotator (${\bf \Omega}\cdot 
{\bf B}< 0$, i.e. pulsar), however, potential difference at the pulsar polar cap 
tends to pull $^{56}$Fe ions from the surface, if the star is a neutron star or a 
strange star with a crust. With $w_{\rm i,NS}\sim 0.9~$keV, one can get 
$T_{\rm cri,i}\simeq (3.5 \times 10^5 {\rm K})B_{12}^{0.73}$ (UM95; UM96).

Given the work function $w$ and $E_\parallel$, another particle ejection mechanism
is the field emission, which is a quantum mechanical tunneling effect and is
only relevant when thermionic ejection is unimportant. Again, by equating the 
current density of the tunneling with the Goldreich-Julian density, one gets the 
critical field to pull certain particles out via field ejection (UM95)
$E_{\rm \parallel,cri}\simeq (6\times 10^{10} {\rm V cm^{-1}}) \left({w\over
1{\rm keV}}\right)^{3/2}$, 
where $w$ is the work function of the particle.

In the RS model, the parallel electric field at surface is $E_\parallel=2\Omega Bh/c$,
where $h$ is the gap height. With (22) of RS75, one gets
$E_{\rm \parallel, CR} \simeq (6.3\times 10^8 {\rm V cm^{-1}}) B_{12}^{3/7} P^{-4/7}$, 
which is smaller than $E_{\rm \parallel,cri}$ for most cases.
It was found that the inverse Compton 
scattering (ICS) induced cascade gaps usually have much smaller gap heights, 
potentials, as well as the surface electric fields (Zhang \& Qiao 1996; Zhang et 
al. 1997; Zhang, Qiao, \& Han 1997). With (12) of Zhang, Qiao \& Han (1997), we get 
even smaller parallel electric fields as
$E_{\rm \parallel, ICS} \simeq (1.4 \times 10^8 {\rm V cm^{-1}}) P^{-2/3}$
for the ``resonant'' ICS induced gaps. Thus, usually field emission is not
important for the charge ejection both for the electrons and ions.

The thermion ejection, however, is important for both cases. Pulsar polar cap
temperature could be estimated self-consistently by the feedback of the inward
particle streams, which is $T=({\gamma m_{\rm e} c^2 \Omega B \over
2 \pi e \sigma})^{1/4}$ by assuming a Goldreich-Julian density,
where $\sigma=5.67 \times 10^{-5} {\rm erg~ cm^{-2}~K^{-4}~ s^{-1}}$ and 
$\gamma$ is the typical Lorentz factor of the primary particles
accelerated within the gap. This turns out to be $T_{\rm CR}\simeq 
(3.1\times 10^6 {\rm K})B_{12}^{3/14}P^{-2/7} $
for CR-induced gap model (RS75), and
$T_{\rm ICS}\simeq (1.5\times 10^6 {\rm K}) P^{-1/3}$
for the ICS-induced gap model (using eq.[14] of Zhang, Qiao \& Han 1997), which 
are much higher than $T_{\rm cri,e}$ and $T_{\rm cri,i}$. Furthermore, hot polar caps
with temperature $>10^6$K have been observed (e.g. Wang \& Halpern 1997). This makes 
the thermionic ejection of both electrons and ions important in most cases and is 
referred as the ``binding energy problem'' of the RS-type vacuum gap model. This
conclusion is more robust for anti-pulsars, and the large error bar of the ion
cohesive energy also only leaves little room for the solution of the binding energy 
problem for pulsars.

The critical temperatures $T_{\rm cri,e}$ and $T_{\rm cri,i}$ could be raised in stronger
magnetic fields. Usov \& Melrose (UM95; UM96) thus developed a modified RS-type model
which operates in strong field pulsars with $B\geq 0.1B_{\rm c}\sim 4.4\times 10^{12}$G, 
where $B_{\rm c}\simeq 4.4\times 10^{13}$G is the critical magnetic field. However, in their 
``self-consistent'' model, the gap is still not completely vacuum, since they invoked 
a partial screening of $E_\parallel$ so as to keep the inflow current not be able to 
heat the surface to above the critical temperatures. Though they didn't 
discuss the ``sparking'' and ``drifting'' process in such model, the results should 
be quite different from the RS's predictions.


\section{PSR 0943+10 as a bare strange star}

Rankin \& Deshpande's work (Rankin \& Deshpande 1998, Deshpande \& Rankin 1999)
showed that, at least
for PSR 0943+10, the drifting sparking patterns just closely match the prediction
of the RS model. In fact, Rankin addressed that the RS model ``is the only
reasonably complete explanation for the hot spots at the moment'' (Glanz 1999).
However, the above-mentioned binding energy problem makes it difficult to
understand the formation of an RS gap in this pulsar if the pulsar is a neutron
star or a strange star with a crust. Even the modified partial screening gap
model proposed by Usov \& Melrose could not heal it since PSR 1943+10 has a 
relative long period of 1.1 second, and a moderate surface field strength
of $2.0\times 10^{12}$G (or $4.0\times 10^{12}$G as Usov \& Melrose argued), which
is well located outside the ``self-consistent'' region of Usov \& Melrose (region 
2 of Fig.1 of UM96). We'll show here that a sound answer could
be obtained if this pulsar is actually a {\em bare strange star} (BSS).

The main objection of BSSs acting as pulsars lies in the superstrong electric 
fields near the star surfaces (Alcock, Farhi, \& Olinto 1986). However, Xu \& 
Qiao (XQ98) showed that the electric field due to the non-neutral effect of
strange stars near a bare strange star's surface actually decreases rapidly. A 
handy calculation using Alcock et al's equation (14) comes to the conclusion 
that the parallel electric field strength will drop from the high value of 
$\sim 5\times 10^{17}{\rm V~ cm^{-1}}$ down to $10^{10} {\rm V~ cm^{-1}}$ 
within a height of $z_{\rm c}\sim 10^{-7}$cm, where rotation-induced electric fields 
begin to dominate (XQ98). Just define an ``effective'' BSS surface at $z_{\rm c}$, 
a magnetosphere like that of a normal neutron star could be formed right above 
this effective surface within a short time scale once some high energy 
$\gamma$-ray seeds can ignite pair production cascade (XQ98). A BSS can hence 
act as a pulsar.

The key advantage of such a BSS model is that a BSS can completely prevent both 
the thermionic and field ejections of any charged particles from the surface.
In other words, the binding energy of the particles in the pulsar surface is 
merely infinity. For the case of ${\bf \Omega}\cdot {\bf B} < 0$ (pulsar 
case), this conclusion is just straight forward, since the positive charges 
within the surface are $u$ quarks rather than ions. The homopolar generated 
strong fields are solely negligible with respect to the strong interaction 
operating between the quarks. Thus essentially $w_{\rm q,BSS}\rightarrow \infty$.
For the case of ${\bf \Omega}\cdot {\bf B} > 0$ (anti-pulsar case) in a BSS,
the situation is a little bit complex. The interaction to prevent electrons 
from ejection is also the electromagnetic force. However, by defining the 
effective surface of BSS at $z_{\rm c}$, the picture could be simplified. At the
effective surface, the homopolar field strength is just equal to the ``binding''
field strength, so that the ``field ejection'' condition fails below it. The
``thermionic ejection'' condition, on the other hand, also fails just
slightly below the effective surface. This is because that, the electric fields 
increase rapidly inwards below $z_{\rm c}$, so that the binding energy of the 
electrons at $z$, $w_{\rm e,BSS}=\int_{z_{\rm c}}^z (dV/dz) dz$ also increases 
rapidly. Note again that the thermionic emission current density is proportional 
to $\exp(-w/kT)$ (RS75, UM95), and that the critical temperature is defined by 
equating this current density with the Goldreich-Julian density, then the critical 
temperature in this BSS case is just proportional to $w_{\rm e,BSS}$ and therefore 
also increases tremendously slightly below the effective surface. As a result, only 
a very thin layer of electrons could be thermionically ejected, and contribute 
a negligible current density, so that a vacuum gap analogous to the RS-type could 
be formed.

There is a little difference between such kind of gap (rooted on a BSS) and 
the original RS-gap (rooted on a NS). The key point is that besides the homopolar
generated electric field, there is also an intrinsic electric field due to the
attraction of the strange quark matters from the BSS. However, the rapidly 
decreasing behavior of this intrinsic or background field (Fig.1 in XQ98) makes 
it play an negligible role. Thus we can safely say that a gap rooted on a BSS can 
completely reproduce all the features of the RS model, which can interpret 
the work of Deshpande \& Rankin (1999) successfully. In this sense, we suggest 
that PSR 0943+10 might be a {\em bare strange star} rather than a normal neutron 
star. As shown above, this argument is 
more promising if it can be inferred from the observations that the star is an 
``anti-pulsar''. We further suggest that other drifting pulsars (e.g. Rankin 1986)
might also be BSSs since most of them have the similar period and surface field
strengths as PSR 0943+10, and the binding energy problem could not be released
if their gaps are rooted on neutron stars or strange stars with crusts.

\section{Conclusion \& Discussion}

We have shown in this {\em Letter} that the lack of theories to solve the
binding energy problem of the RS model rooted on a neutron star has led us to
present the idea that PSR 0943+10 as well as other clearly drifting pulsars
might be bare strange stars. Though the argument in favor of the BSS model
is indirect, it seems that this is the only hitherto known sound model to 
solve the binding energy problem completely.

There are some arguments against the formation of the BSSs or
even strange stars. In their pioneer paper, Alcock, Farhi \& Olinto (1986)
simply addressed that ``a bare strange star may readily accrete some ambient
material'' since ``the universe is a dirty environment''. Though firmly
concluding that all accreting X-ray pulsars have crusts, they admitted 
however that ``the situation with radio pulsars is harder to assess'', and
that ``the rotating magnetosphere is likely to prevent fluid accretion''.
Thus as long as the fallback materials do not form a crust during the
supernova explosion when a strange star were born, 
the BSS could remain bare for a sufficient long
period of time before its rotation is too slow to prevent materials
to drop onto its polar cap. At this stage, usually pulsars have died out
across the ``death lines'' or ``death valleys'' (RS75; Chen \& Ruderman 1993;
Qiao \& Zhang 1996). Thus it is plausible to say that observed pulsars could
be BSSs.

Perhaps the most severe argument against the existence of strange stars
is the ``glitching'' phenomena observed from some pulsars, which is 
commonly interpreted as the starquakes happened in a solid crust. Even 
for strange stars with solid crusts, Alpar (1987) argued that the observed 
magnitude of $\Delta\dot\Omega /\dot\Omega \sim 10^{-2}-10^{-3}$ poses a 
strong objection to such an idea, since strange stars' crusts are not thick 
enough. These arguments are not in conflict with our idea that drifting 
pulsars might be BSSs. By comparing the samples of the drifting pulsars 
(Table 2 of Rankin 1986) and the samples of the glitching pulsars (Table 
6.2 of Lyne \& Graham-Smith 1998), we found remarkably that, all other 
drifting pulsars were never observed to show glitching behavior except for 
PSR 0525+21, which is one of the few pulsars with surface magnetic field 
higher than $10^{13}$G. In such a high field, the gap of the UM96's type 
or even RS75's type could be formed in a {\em neutron star} surface, since 
the binding energies are greatly enhanced. Furthermore, glitching behavior 
might also be interpreted by the BSS models, if stable, low-baryon number 
strangelets could exist and form a solid crust (Benvenuto \& Horvath 1990). 
Thus observed glitches ``should not be used to dismiss the possibility of 
strange stars'' (Madsen 1998).

Another possible objection of our idea may come from strong thermal X-ray
emissions from drifting pulsars, since the bare quark matter surface of a
strange star is a very poor radiator itself. Spectral analysis of some 
spin-powered X-ray pulsars (e.g. Becker \& Tr\"umper 1997) reveals that 
the X-ray spectra can be fitted by either a power-law radiation (non-thermal 
magnetospheric origin), a thermal emission from the full surface (mainly due 
to the cooling or the internal heating), a thermal emission from the hot 
polar cap (due to inner gap or outer gap heating), or a combination of above 
two or three components. Our model actually predicts that {\em the full surface 
thermal emission from the drifting pulsars should be strongly suppressed}. 
Four pulsars, i.e., Vela, Geminga, PSR 0656+16, and PSR 1055-52, are observed 
to show strong thermal emission from the full surface. But they are not 
drifting pulsars, and thus can not dismiss our idea. Future X-ray observation 
and spectral analysis on drifting pulsars can prove or dismiss the idea that 
drifting pulsars are BSSs.

The final question is whether a strange star could be formed in the supernova 
explosion at all. No definite answer is available yet. Nevertheless, as argued 
in XQ98, the birth of a strange star rather than a neutron star could enhance 
both the successful possibilities of supernova explosion and the energy of the 
revived shock wave, due to the additional energy source of phase transition from
two-flavor quark matter to three-flavor quark matter (Dai, Peng, \& Lu 1995).

In principle, a BSS model can mend the RS vacuum gap model to have a much more
solid foundation. The existence of such a gap can benefit the inverse Compton 
scattering model of pulsars (Qiao \& Lin 1998), which can interpret naturally 
the long identified pulsar ``core'' emission (Rankin 1983; Lyne \& Manchester 
1988).

The idea presented here also adds one more criterion to distinguish strange
stars from neutron stars. As shown above, the arguments in favor of the BSSs
are more promising for the anti-pulsar case. Unfortunately, the drifting
direction does not depend on whether the star is a pulsar or an anti-pulsar.
Thus seeking other observational methods to tell the sense of the magnetic 
pole of a pulsar is essential. We propose here that {\em finding and 
identifying a drifting anti-pulsar will be a strong argument in favor of 
the existence of the (bare) strange stars in nature}.

\begin{acknowledgments}

We sincerely thank Joanna Rankin for discussing her recent work (with Deshpande)
with us, the anonymous referee for insightful suggestions, 
and M. Vivekanand for valuable comments.
Helpful discussions
from J.L. Han and other members of our pulsar group are acknowledged.
This work is supported by the National Natural Science Foundations (No.19673001
and No. 19803001)
of China, the Climbing Program of State Committee of Science and Technology of 
China, the Research Fund for Doctoral
 Program 
of Higher Education, and the 
Youth Science Foundation of Peking University.

\end{acknowledgments}

\end{document}